\documentclass[a4paper]{revtex4}
\usepackage{graphicx}
\usepackage{fancyhdr}
\usepackage{amsmath}
\usepackage{color}

\begin{document}

\title{Type-II seesaw model with two-Higgs doublets}

%

\author{Takaaki Nomura}
\affiliation{Department of Physics, National Cheng-Kung University, Tainan 701, Taiwan  }

\begin{abstract}
 A study of searching for doubly charged Higgs $(\delta^{\pm\pm})$ is performed in two-Higgs-doublet extension of the type-II seesaw model. We find that a significant mixing effect between singly charged Higgs of Higgs doublet and of triplet is arisen from the scalar potential. The mixing leads to new production processes and decay patterns of doubly charged Higgs. With luminosity of 40 fb$^{-1}$ and collision energy of 13 TeV at the LHC, we show that $\delta^{\pm\pm}$ with mass below $330$ GeV could be observed at the $5\sigma$ level. Moreover,  for the luminosity of 300 fb$^{-1}$, the observed mass of $\delta^{\pm\pm}$ could reach up to 450 GeV. 
 \end{abstract}

\maketitle

\thispagestyle{fancy}


\section{Introduction}
The origin of neutrino mass is one of the unsolved issue in the standard model (SM).
This mystery could be solved by a Higgs triplet extension of the SM called Type-II seesaw model~\cite{Magg:1980ut, Konetschny:1977bn}.
The novel feature of a Type-II seesaw model is the existence of a doubly charged Higgs $\delta^{\pm \pm}$ in the Higgs triplet $\Delta$.
In the original Type-II seesaw model, $\delta^{\pm \pm}$ decays into two same-sign charged leptons $\ell^\pm \ell'^\pm$ or W bosons $W^\pm W^\pm$ according to the triplet vacuum expectation value (VEV)
when masses in the multiplet are degenerate.
Thus the experimental searches of doubly charged Higgs bosons have focused on the signals of same-sign dilepton or W bosons~\cite{Chatrchyan:2012ya, ATLAS:2012hi, Khachatryan:2014sta}. 

The extension of Type-II seesaw model by including second Higgs doublet can lead new effect in doubly charged Higgs production and decays~\cite{Chen:2014xva, Chen:2014qda}.
The most attractive new effect is the dimension-3 terms in scalar potential, read by $\mu_x H^T_j i\tau_2 \Delta^\dagger H_k$ (j,k =1,2; $x$=1,2,3), where $H_j$ is a Higgs doublet. Since the coefficients $\mu_x$ can be of order of electroweak (EW) scale, the new terms lead a large mixing angle  between the singly charged Higgs of doublet ($H^\pm$) and of triplet ($\delta^{\pm}$).
Consequently the doubly charged Higgs can be produced via QCD process owing to the the mixing effect.

In this study we focus on the new production and decay patterns of $\delta^{\pm \pm}$.
We then carry out a simulation study for the signal of $\delta^{\pm \pm}$ at the LHC and estimate the significance of discovering the doubly charged Higgs.

\section{The interactions of doubly charged Higgs boson}
To investigate the production and decay of doubly charged Higgs boson $\delta^{\pm \pm}$, we first discuss the relevant interactions of  charged Higgs bosons in the two Higgs doublet(THD)-Type-II seesaw model.
The gauge interactions and Yukawa couplings of $\Delta$ and $H_{1,2}$ are same as original Type-II seesaw model and THD model (Type-II) respectively. We thus do not explicitly write down them here; more details are given in Refs~\cite{Chen:2014xva, Chen:2014qda}. 
The Higgs potential is written such that
\begin{align}
 V(H_1, H_2, \Delta) &=  V_{H_1H_2} + V_\Delta + V_{H_1 H_2 \Delta}\,, \nonumber \\
V_{H_1H_2} &= m^2_1 H^\dagger_1 H_1 + m^2_2 H^\dagger_2 H_2 - m^2_{12} ( H^\dagger_1 H_2 + h.c.)+ \lambda_1 ( H^\dagger_1 H_1)^2  \nonumber \\
 &+  \lambda_2 (H^\dagger_2 H_2)^2 + \lambda_3  H^\dagger_1 H_1 H^\dagger_2 H_2+ \lambda_4  H^\dagger_1 H_2 H^\dagger_2 H_1 + \frac{\lambda_5}{2} [(H^\dagger_1 H_2)^2+h.c. ] \,, \nonumber \\
 V_\Delta &= m^2_\Delta Tr \Delta^\dagger \Delta + \lambda_{9} (Tr \Delta^\dagger \Delta)^2 + \lambda_{10} Tr (\Delta^\dagger \Delta)^2\,, \nonumber \\
 V_{H_1H_2\Delta} &= \left( \mu_1 H^T_1 i\tau_2 \Delta^{\dagger}  H_1 + \mu_2 H^T_2 i \tau_2 \Delta^\dagger H_2 + \mu_3 H^T_1 i\tau_2 \Delta^\dagger H_2 + h.c. \right) \nonumber \\
 &+ ( \lambda_6 H^\dagger_1 H_1 + \bar \lambda_6 H^\dagger_2 H_2 ) Tr \Delta^\dagger \Delta + H^\dagger_1 ( \lambda_7 \Delta \Delta^\dagger + \lambda_8 \Delta^\dagger \Delta ) H_1 
+  H^\dagger_2 \left( \bar \lambda_7 \Delta \Delta^\dagger + \bar \lambda_8 \Delta^\dagger \Delta \right) H_2\,, \label{eq:v}
  \end{align}
 where $V_{H_1 H_2}$ and  $V_\Delta$ denote the scalar potential of THD and of pure triplet, and  $V_{H_1 H_2 \Delta}$ is the part involving $H_1$, $H_2$ and $\Delta$.  The VEV of the triplet $v_\Delta$ is required to be 
  \begin{equation}
  v_\Delta \approx \frac{1}{\sqrt{2}} \frac{\mu_1 v^2_1 + \mu_2 v^2_2 + \mu_3 v_1 v_2}{m^2_\Delta + (\lambda_6+\lambda_7) v^2_1/2 + (\bar\lambda_6+\bar\lambda_7)v^2_2 /2}\,, \label{eq:v_d}
  \end{equation}
  where $v_{1,2}$ is the VEV of $H_{1,2}$. 
  We only keep the leading power for $v_\Delta$ in Eq.~(\ref{eq:v_d}) since it should satisfy $v_\Delta << v=\sqrt{v_1^2 + v_2^2}$ from $\rho$ parameter measurement. 
  Interestingly, the parameters $\mu_{1,2,3}$ as large as electroweak scale can provide $v_\Delta << v$ if the parameters satisfy the relation $\mu_3 \sim - (\mu_1 v_1^2 + \mu_2 v_2^2)/(v_1 v_2)$ in contrast to original Type-II seesaw model. In our analysis, we adopt the relation for simplicity.
  
Two physical singly charged Higgs bosons are provided by both THD sector and Higgs triplet which are denoted by $H^\pm$ and $\delta^\pm$ respectively.
The mass eigenstates are obtained as a combination of $H^\pm$ and $\delta^\pm$ as
\begin{equation}
  \left( \begin{array}{c}
    H^\pm_1\\ 
    H^\pm_2 \\ 
  \end{array}\right) =   \left(\begin{array}{cc}
    \cos\theta_\pm & \sin\theta_\pm \\ 
    -\sin\theta_\pm & \cos\theta_\pm\\ 
  \end{array}\right) \left( \begin{array}{c}
    H^\pm \\ 
    \delta^{\pm} \\ 
  \end{array}\right)\,.\label{eq:ma}
 \end{equation}
  The masses and their mixing angles are written by
   \begin{align}
 \left( m_{H_{1,2}^{\pm}}\right)^2 =  \frac{1}{2}\left(m^2_{\delta^\pm} + m^2_{H^\pm}\right) \mp \frac{1}{2} \left[ \left( m^2_{\delta^\pm} - m^2_{H^\pm}\right)^2 + 4 m^4_{H^-\delta^+}\right]^{1/2}\,, \quad
 \tan2\theta_\pm = - \frac{2 m^2_{H^- \delta^+}}{m^2_{\delta^\pm} - m^2_{H^\pm}}\,,
 \label{eq:mass_mixing}
  \end{align}
 where $H_1^\pm$ is identified as the lighter charged Higgs, $m^2_{H^\pm} = (m_{12}^2 - (\lambda_4 + \lambda_5)v_1 v_2/2)/(\sin \beta \cos \beta)$, $m^2_{H^- \delta^+}=v[\mu_1 \cos^4 \beta -\mu_2 \sin^4 \beta +(\mu_1 -\mu_2)\sin^2 \beta \cos^2 \beta]/(2 \sin \beta \cos \beta) $, and $m^2_{\delta^\pm} = m_\Delta^2 +v_1^2(2 \lambda_6 + \lambda_7 + \lambda_8)/2 + v_2^2(2 \bar \lambda_6 + \bar \lambda_7 \bar \lambda_8)/2$.
 The magnitude of the mixing angle $\theta_\pm$ depends on the massive parameter $\mu_{1,2}$. We are going to explore the influence of a large $\theta_\pm$ on the search of doubly charged Higgs.
 
 In our model, two singly charged Higgses can interact with quarks owing to the large mixing effect.
 The $H_{1,2}^\pm$-quark interactions are given by
 \begin{align}
\frac{\sqrt{2}}{v}  \left[\bar u\left( \tan\beta V_{CKM} {\bf m_D}   P_R + \cot\beta  {\bf m_U} V^\dagger_{CKM} P_L \right) d  \right]
(\cos\theta_\pm H^+_1 -  \sin\theta_\pm H^+_2) + h.c. \,, \label{eq:hff}
  \end{align}
   where we suppress all flavor indices, $u^T=( u, c, t)$ and $d^T = ( d, s, b)$ denote the up and down type quarks, $V_{CKM}$ is the Cabibbo-Kobayashi-Maskawa (CKM) matrix, ${\bf m_{D(U)}}$ is the diagonalized mass matrix of down (up) type quarks, and $P_{R, L} = (1 \pm \gamma_5)/2$. 
   
\section{Production and decays of doubly charged Higgs boson}

The doubly charged Higgs could be produced by EW interactions via the s channel processes:
 \begin{align}
& p p \rightarrow Z/\gamma \rightarrow \delta^{++} \delta^{--} \,, \label{eq:EW1}\\
& p p \rightarrow W^\pm \rightarrow \delta^{\pm \pm} H_{1,2}^{\mp} \,. \label{eq:EW2}
\end{align}
Except for the new mixing effect $\theta_+$, the production channels are similar to those in the original Type-II seesaw model.
Moreover, with the sizable mixing angle $\theta_+$, the on-shell $\delta^{\pm \pm}$ could be produced through QCD interactions:
\begin{align}
 \label{eq:QCDI}
 & p p \rightarrow H_2^+ \bar{t} b \ (H_2^- t \bar b) \rightarrow \delta^{++} W^- \bar{t} b \ (\delta^{--} W^+ t \bar b) \,,  \\
 \label{eq:QCDII}
& p p \rightarrow H_2^+ \bar{t} \ (H_2^- t)  \rightarrow \delta^{++} W^- \bar{t} \ (\delta^{--} W^+ t)\,. 
 \end{align}
The on-shell $\delta^{\pm \pm}$ in  Eq.~(\ref{eq:QCDI}) and (\ref{eq:QCDII}) is generated by the decay $H^\pm_2 \to \delta^{\pm\pm} W^{\mp}$ since 
we adopt mass relation of $m_{H^\pm_1} < m_{\delta^{\pm \pm}} < m_{H^\pm_2}$.
 The production of $\delta^{\pm\pm}$ through lighter charged Higgs $H^\pm_1$ is  off-shell  effects and small, we therefore ignore its contributions. For the processes in Eq.~(\ref{eq:QCDII}), the main QCD reaction is associated with the interactions of b-quark and gluons, e.g. $\bar b(b) g \to H^+_2 \bar t ( H^-_2 t)$.

In our analysis, we apply following conditions for the parameters in the model:
\begin{align}
m_{\delta^\pm} = m_{\delta^{\pm \pm}} + 100 \ {\rm GeV} \,, \quad
m_{H^\pm} = \frac{4}{5} m_{\delta^{\pm}} \,, \quad
\mu_1 = -\mu_2 = m_{\delta^\pm} \sin \beta \cos \beta \,, \label{eq:setting}
\end{align}
which give sizable mixing angle $\theta_+$ and the mass relation of $m_{H^\pm_1} < m_{\delta^{\pm \pm}} < m_{H^\pm_2}$.
The masses of singly charged Higgs and mixing angle are then obtained from Eq.~(\ref{eq:mass_mixing}).
 
\begin{figure}[t]
\centering
\includegraphics[width=70mm]{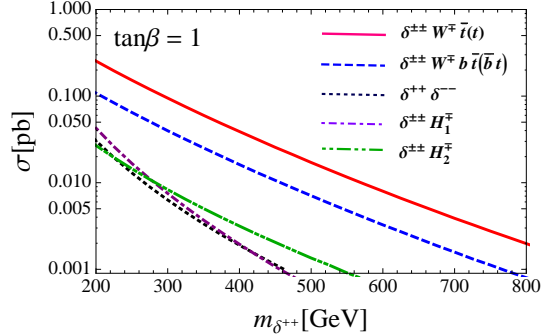}
\caption{Product cross sections of the doubly charged Higgs as a function of mass for a collision energy of 13 TeV at the LHC; the dotted, dash-dotted, and dash-dot-dotted lines denote the EW processes, while the solid and dashed lines stand for the QCD processes. The parameter setting of Eq.~(\ref{eq:setting}) and $\tan \beta =1$ are used. (reproduced from 
~\cite{Chen:2014qda}.)} \label{fig:XS}
\end{figure}

 We then calculate the cross sections of $\delta^{\pm \pm}$ production processes using {\it CalcHEP 3.6.15} code~\cite{CalcHEP} by implementing the model.
Applying the settings of Eq.~(\ref{eq:setting}) and $\tan\beta =1$, we show the production cross sections for the processes  in Eqs.~(\ref{eq:EW1})-(\ref{eq:QCDII})  as a function of $m_{\delta^{\pm\pm}}$ in Fig.~\ref{fig:XS},  where  the collision energy at LHC is 13 TeV and {\tt CTEQ6L} PDF is used; the dotted, dash-dotted and dash-dot-dotted lines denote the EW processes while the solid and dashed lines stand for QCD processes, respectively. 
We therefore find that QCD production cross sections are significantly larger than that of EW production in our setting.

The produced $\delta^{\pm \pm}$ can decay into $\ell^\pm \ell^\pm$, $W^\pm W^\pm$, $H^\pm_i H^\pm_j$, $W^\pm H^\pm_j$, etc.
With our parameter setting, the dominant decay modes are $W^\pm H^\pm_{1} (H^{*\pm}_2)$ where $H^{*\pm}_2$ is off-shell, 
since we assume $v_\Delta << v$ and small Yukawa couplings for triplet suppressing $W^\pm W^\pm$ and $\ell^+ \ell^+$ modes. 
Then the lighter singly charged Higgs $H^{+(-)}_{1}$ dominantly decays into $t \bar b(\bar t b)$; the off-shell  $H^{*+(-)}_2$ also coverts to $t \bar b(\bar t b)$.

\section{Simulation study}

In this section, we discuss the simulation study of possible signal and background events.
We then estimate the significance of discovering the doubly charged Higgs after applying relevant kinematic cuts (see also \cite{Chen:2014qda} for details).
In order to generate the simulation events, we employ the {\tt MADGRAPH/MADEVENT\,5}~\cite{Ref:MG} and use {\tt PYTHIA\,6}~\cite{Ref:Pythia} to deal with the fragmentation of hadronic effects,  the  initial-state radiation (ISR) and final-state radiation (FSR) effects, and the decays of SM particles e.g. $W$-boson, $t$-quark, etc.
In addition, the generated  events are also run though the {\tt PGS\,4} detector simulation~\cite{Ref:PGS}.  

The $\delta^{\pm \pm}$ is dominantly produced by the QCD production processes Eqs.~(\ref{eq:QCDI}) and (\ref{eq:QCDII}), 
and decays as $\delta^{++(--)} \to W^{+(-)} H^{+(-)}_{1,2} \to W^{+(-)} t \bar b (\bar t b) \to W^{+(-)} W^{+(-)} b \bar b$.
We thus require the signal events as 
\begin{equation}
\ell^\pm \ell^\pm + {\rm four \, or \, more \, jets}\,.  \label{eq:sig}
\end{equation} 
For background processes we consider following final states at the LHC:
$\ell^+ \ell^-$ (+ISR/FSR),  $W^\pm W^\pm j j (\alpha^4)$, $W^\pm W^\pm j j (\alpha^2 \alpha_s^2)$, $W^\pm t \bar{t}$, $W^\pm t \bar{t} j$,   
$W^\pm Z +n j$ and $Z Z+ nj$, 
where the number of jets $n$ for VV backgrounds is taken as $n \leq 2$. $W^\pm W^\pm +nj$ events in $VV$ background have been included in EW and QCD background, therefore they should be excluded. 

For enhancing the significance of $\delta^{\pm\pm}$ signals by reducing the possible backgrounds, we need to propose some strategies of kinematical cuts. For excluding the soft leptons and jets, when we generate the events by event generator, 
%
we set the preselection conditions for leptons and jets to be 
\begin{align}
\label{eq:cuts_basic}
p_T (\ell) > 10 \ {\rm GeV}, \quad \eta(\ell) < 2.5, \quad p_T (j) > 20 \ {\rm GeV}, \quad \eta(j) < 5.0, 
\end{align}
where $p_T$ is the transverse momentum and $\eta = 1/2 \ln (\tan \theta/2)$ is pseudo-rapidity  with $\theta$ being the scattering angle  in the laboratory frame.
Furthermore we apply the selection cuts
\begin{align}
N_{\rm b-jet} \geq 1, \quad p_T(\ell_2) < 60 \ {\rm GeV}\,, \quad
M_{\ell^\pm \ell^\pm} < \frac{m_{\delta^{\pm \pm}}}{4}. \label{eq:Mll}
\end{align}
where $N_{\rm b-jet}$ denotes the number of b-jet, $\ell_2$ stands for the second highest $p_T$ charged lepton.

\begin{figure}[t] 
\includegraphics[scale=0.7]{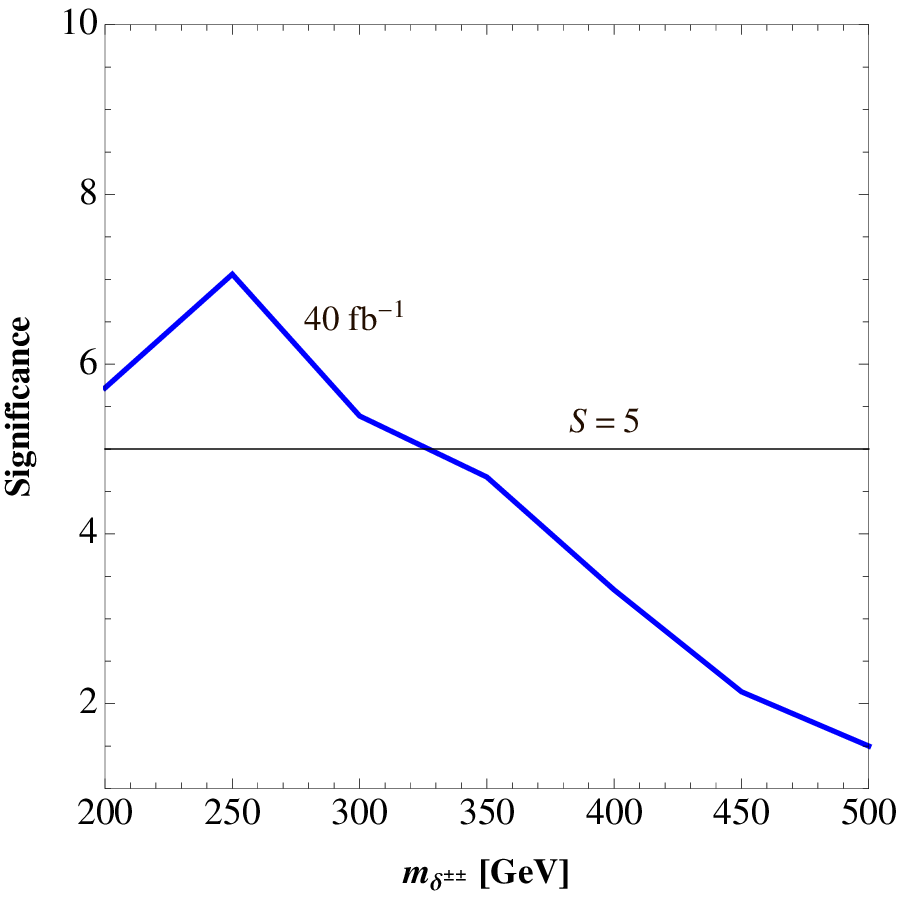} 
\includegraphics[scale=0.7]{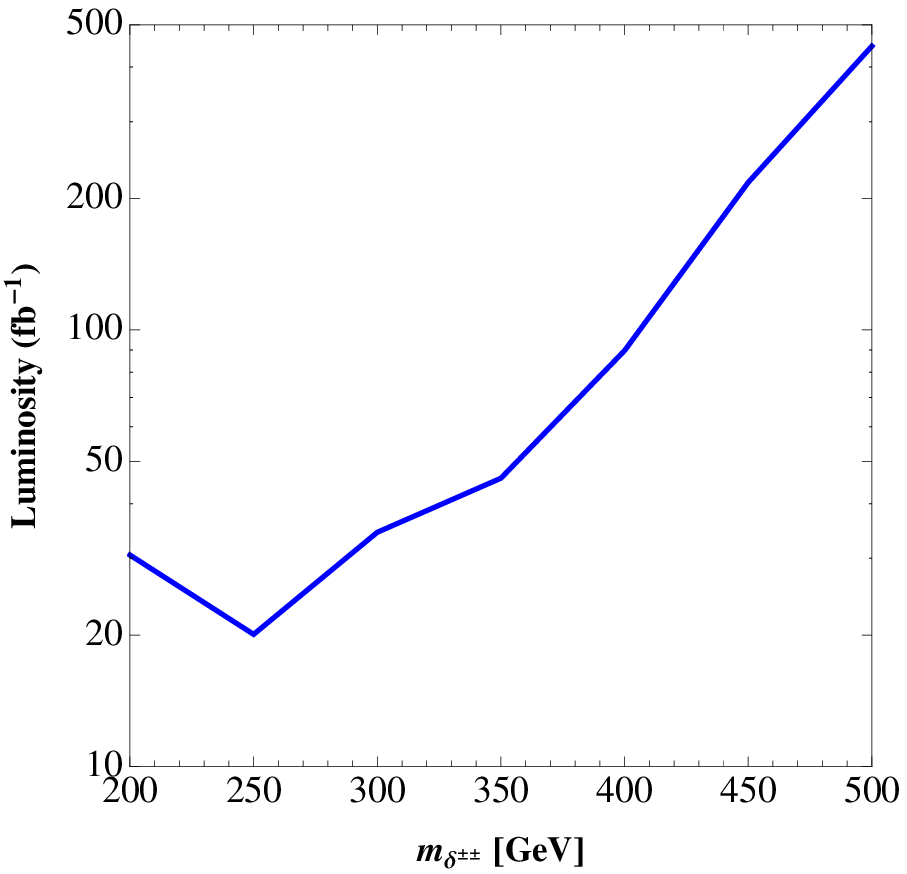} 
\caption{  Significance with 40 fb$^{-1}$ (left) and luminosity for $S > 5$ (right) as a function of $m_{\delta^{\pm\pm}}$.  The collision energy of 13 TeV is applied for both plots.  (reproduced from 
~\cite{Chen:2014qda}.)
\label{fig:significance}}
\end{figure}

Finally we estimate significance which is defined as~\cite{Ball:2007zza} 
$S = \sqrt{2[(n_s+n_b)\ln (1+n_s/n_b)-n_s]}$,
where $n_s$ and $n_b$ denote the number of signal and background events, respectively.
The left (right) panel of Fig.~\ref{fig:significance} is the estimated significance (luminosity to get 5$\sigma$ discovery) as a function of $m_{\delta^{\pm\pm}}$. By the figure, 
one can find that  the doubly charged Higgs with a mass lower than 330 GeV can be discovered at the LHC with an integrated luminosity of 40 fb$^{-1}$. Additionally,  the doubly charged Higgs with a mass of 450 GeV can be discovered at the LHC with an integrated luminosity of 300 fb$^{-1}$.

%

\section{Summary}
We have discussed the Type-II seesaw model with two Higgs doublet.
The new interaction in the model can lead the significant mixing effect between singly charged Higgs of the Higgs doublet and triplet.
The mixing results in new production processes and decay patterns of the doubly charged Higgs.

We have shown that QCD processes  are the predominant effects to produce the $\delta^{\pm\pm}$, read as $pp \to H_2^+ \bar t b (H_2^- t \bar b) \to \delta^{++} W^- \bar t b (\delta^{--} W^+ t \bar b)$ and $\bar b(b) g \to H_2^+ \bar t (H_2^- t) \to \delta^{++}W^- \bar t (\delta^{--} W^+ t)$ owing to the mixing effect, while other Higgs triplet models  are arisen from EW processes. 
Subsequently the doubly charged Higgs decays as $\delta^{++(--)} \to W^{+(-)} H^{+(-)}_{1,2} \to W^{+(-)} t \bar b (\bar t b) \to W^{+(-)} W^{+(-)} b \bar b$.

We then have investigated the significance of doubly charged Higgs signal by numerical simulation including relevant kinematical cuts.
We find that with luminosity of 40 fb$^{-1}$ and collision energy of 13 TeV, $\delta^{\pm\pm}$ with mass below $330$ GeV could be observed at the $5\sigma$ level. Additionally, the observed mass of $\delta^{\pm\pm}$ could be up to 450 GeV when the luminosity approaches 300 fb$^{-1}$.

\begin{acknowledgments}
 This work is supported by the Ministry of Science and Technology of 
R.O.C. under Grant \#: MOST-103-2811-M-006-030. We also thank the National Center for Theoretical Sciences (NCTS) for supporting the useful facilities. 
\end{acknowledgments}

\bigskip 

\end{document}